# On the topological nature of Volterra's theorem

**D. H. Delphenich**[†]
Kettering, OH, USA 45440

**Key words** Strain compatibility, Volterra construction, dislocations and disclinations

**PACS** 02.40.Hw, 02.40.Re, 02.40.Yy, 46.25.Cc

It is first observed that the original formulation of the Volterra construction for dislocations and disclinations was related to the role that homotopy plays in strain compatibility, whereas the modern discussions are chiefly concerned with how it relates to the holonomy groups of connections that have non-vanishing torsion and curvature. However, the Saint Venant conditions that follow from assuming infinitesimal strain compatibility imply that both torsion and curvature must vanish. The resolution of the confusion is in the fact that when a manifold is multiply connected a flat connection might still have non-trivial discrete holonomy.

## 1 Introduction

Nowadays, the accepted way of presenting the so-called "Volterra construction" of dislocations and disclinations in crystal lattices (see, e.g., [**1-3**]) is purely focused on the *geometrical* aspects of what Volterra was discussing in his seminal paper [**4**] of 1907 One basically parallel-translates a tangent vector around a loop in the lattice by means of a somewhat vaguely-defined connection and concludes, by holonomy, that a resulting non-vanishing Burger vector implies non-vanishing torsion, while a non-vanishing Frank vector implies non-vanishing curvature.

However, if one goes back to Volterra's paper then the first thing that one notices is that the title "Sur l'équilibre des corps élastiques multiplement connexes" has more to do with the role of *topology* – in the form of homotopy – than it does with geometry, at least directly.

One finds that Volterra's motivation for the main theorem of the treatise was the previously-established fact that when an irrotational, incompressible fluid occupies a simply-connected region, its equilibrium state – i.e., the state of minimum kinetic energy – is the rest state uniquely, but when it occupies a multiply-connected region it is possible for the equilibrium state to have non-vanishing kinetic energy. Indeed, one can now cite the example of superfluids moving in toral regions as a laboratory example of how this is manifested in nature.

The main theorem of the Volterra treatise was then the extension of that result to the deformations of elastic media. In particular, he showed that the equilibrium state of a simply-connected elastic body was a state of zero strain, while the equilibrium state of a multiply-connected one could very well have non-vanishing strain. One should observe that his proof was therefore not one of *existence*, rigorously speaking, but of *possibility*. It was in the process of exhibiting that possibility that he introduced the well-known Volterra construction of a multiply-connected elastic body by the identification of the end faces of a solid cylinder by means of rigid motions of them relative to each other.

[†] E-mail: david_delphenich@yahoo.com

As an example of his theorem, one sees that if one deforms a rubber rod in a homotopic sort of way and releases it then it will return to its undeformed, unstrained state in equilibrium, but if one also attaches the end faces to each other in some way then when one releases it the resulting configuration the equilibrium state will be deformed and strained. It is essential to see that such a deformation cannot come about by a homotopy alone, since the initial rod is not homotopically equivalent to the final solid torus; in particular, the fundamental group has acquired a single non-trivial generator. More to the point, the initial state of deformation is not diffeomorphic to the final state.

Since Volterra did not introduce a connection or demand parallel translation at any point in his proof – at least explicitly – one then desires to see how the purely homotopic proof that he presented eventually gave way to the essentially geometric explanation that one now finds to be customary. The resolution of the confusion is actually quite straightforward: When a manifold is multiply connected, an affine connection can have non-trivial holonomy while still having vanishing torsion and curvature.

Of course, this fact is already well-known to differential geometry [**5-7**], but in the present context of continuum mechanics it carries with it a subtlety: The fundamental issue in Volterra's theorem is that of the integrability of a given strain into a deformation that caused it. As we pointed out, if a deformation is to take the form of a diffeomorphism of an initial state of deformation then clearly if the final state is not homotopically equivalent to the initial one then no such deformation can exist. Since the Saint Venant integrability conditions for strain can be expressed as saying that the deformed metric on the final state must still have vanishing curvature one sees that the question of integrability carries with it both topological and geometric aspects. Indeed, one sees that, in effect, the "topological defect" that the dislocation or disclination represents is serving as the "source" of non-trivial geometry, at least in the eyes of holonomy.

The statement and proof of Volterra's theorem consisted of two general phases, namely, examining the strain compatibility conditions for multiply connected elastic bodies, and then applying the results to the study of the equilibrium state. Because the main idea that we are going to discuss in the present study is concerned with how topology and geometry relate to strain compatibility, we shall only address the first part of his theorem.

The basic direction of the present discussion of that theorem is then as follows: In the next Section, we first present the theory of strain in a modern geometrical fashion, including both the Volterra derivation of the strain compatibility conditions, using an alternate form that does not suffer from the same limitations as his did. In Section 3, we then elaborate on the nature of the connection that was implicitly introduced in the proof of strain compatibility, namely, the teleparallelism connection for a certain coframe field, and how it relates to the connection that is introduced in deriving the Saint Venant conditions. We then recall some of the known facts about the geometry of flat manifolds as they relate to multiple-connectedness, and finally summarize the main results and discuss how they relate to other established topics of modern mathematical physics, such as topological defects in ordered media and non-trivial vacua in gauge field theories.

## 2 The geometry of strain

The Cauchy-Green concept of strain as a measure of the deformation of a material object $\mathcal{O}$ that has been embedded in some space $M$ is manifestly geometric in character. What one measures is the extent to which the deformation is not an isometry of the ambient metric $g$ on the manifold $M$.

### 2.1 Finite and infinitesimal strain tensors

First, we define an *object* $\mathcal{O}$ to be a compact submanifold with boundary $\partial\mathcal{O}$ in some parameter space $\mathbf{R}^p$, whose coordinates will be expressed in the form $a^i$, $i = 1, \ldots, p$. An *embedding* of $\mathcal{O}$ in a differentiable manifold $M$ is a map $x$: $\mathcal{O} \to M$, $a \mapsto x(a)$ that is a diffeomorphism onto its image $x(\mathcal{O})$. In some coordinate chart $(U, x^\mu)$, $\mu = 1, \ldots, n$ on the $x(\mathcal{O})$, the embedding will be expressed by the systems of equations:

$$x^\mu = x^\mu(a^i). \tag{2.1}$$

By assumption, the differential map $dx|_a : T_a\mathcal{O} \to T_{x(a)}M$ will be a linear map of rank $p$. Relative to the local coordinates we have chosen, it will look like:

$$dx|_a = x^\mu_{,i}(a)\, da^i \otimes \partial_\mu \,. \tag{2.2}$$

If $\bar{x} : \mathcal{O} \to M$ is another embedding of $\mathcal{O}$ then the *finite deformation* of $x$ to $\bar{x}$ is the diffeomorphism $f$: $x(\mathcal{O}) \to \bar{x}(\mathcal{O})$, that is defined by the composition of $x^{-1}$ followed by $\bar{x}$:

$$y = \bar{x} \cdot x^{-1}. \tag{2.3}$$

One refers to $x(\mathcal{O})$ as the *initial state* and $\bar{x}(\mathcal{O})$ as the *deformed state* of $\mathcal{O}$ in $M$.

If $(V, y^\mu)$ is a coordinate chart on $\bar{x}(\mathcal{O})$ that contains $y(U)$ then this deformation can be expressed in the form:

$$y^\mu = y^\mu(x^\nu). \tag{2.4}$$

However, unless $p = n$ this system will not be invertible, in general, and one usually has to use coordinate charts that are adapted to both embeddings in order to obtain an invertible system of equations for the deformation by eliminating the "normal" coordinates from the system by setting them equal to zero for points of $x(\mathcal{O})$ and $\bar{x}(\mathcal{O})$.

Hence, we assume that the aforementioned system (2.4) represents that reduced system, for brevity, and the indices range from 1 to *p*.

Since the deformation is a diffeomorphism of a *p*-dimensional manifold to another one, the differential map $dy|_x : T_x(x(\mathcal{O})) \to T_{y(x)}(\bar{x}(\mathcal{O}))$ must be invertible, so the matrix:

$$y^{\mu}_{,\nu} = \frac{\partial y^{\mu}}{\partial x^{\nu}} \tag{2.5}$$

must be invertible for all $x \in x(\mathcal{O})$.

If the ambient manifold *M* of the embeddings carries a metric *g* then one can pull the metric on $\bar{x}(\mathcal{O})$ back to a metric $y^*g$ on $x(\mathcal{O})$. If **v**, **w** are tangent vector fields on $x(\mathcal{O})$ then this pull-back is defined by:

$$y^* g_x(\mathbf{v}, \mathbf{w}) = g_{y(x)}(dy|_x(\mathbf{v}), dy|_x(\mathbf{w})). \tag{2.6}$$

In local coordinates, if we denote the components of $y^*g$ by $\bar{g}_{\mu\nu}(x)$ then the components of the pulled-back metric are:

$$\bar{g}_{\mu\nu}(x) = g_{\kappa\lambda}(y(x))\, y^{\kappa}_{,\mu}(x) y^{\lambda}_{,\nu}(x)\,. \tag{2.7}$$

One must note the fact that the components of $y^*g$ at $x \in x(\mathcal{O})$ depend upon the components of *g* at $y(x) \in \bar{x}(\mathcal{O})$, which means that when *g* is not a flat metric one might be introducing a potential source of error in the definition by comparing tangent objects at finitely-spaced points of *M* without introducing a connection and addressing the breakdown of parallel translation, which is usually applicable in an unambiguous way only locally. Although this might be an issue in general relativistic continuum mechanics, such as the deformation of objects near neutron stars and black holes, nevertheless, in conventional mechanical engineering, which is quite non-relativistic, one usually lets *M* be Euclidian $\mathbf{R}^3$, whose Levi-Cività connection is flat, so one can find orthonormal coordinate frame fields that make $g_{\mu\nu} = \delta_{\mu\nu}$.

While ignoring these matters of geometrical rigor, we proceed to define the *Cauchy-Green finite strain tensor E* on $x(\mathcal{O})$ to be the difference between $\bar{g}$ and *g:*

$$E = y^*g - g\,. \tag{2.8}$$

One then sees that locally one has:

$$E_{\mu\nu}(x) = \bar{g}_{\mu\nu}(x) - g_{\mu\nu}(x) = g_{\kappa\lambda}(y(x))y^{\kappa}_{,\mu}(x)y^{\lambda}_{,\nu}(x) - g_{\mu\nu}(x). \tag{2.9}$$

In conventional continuum mechanics [**8-11**], which takes place in Euclidian $\mathbf{R}^3$, it is traditional to introduce the *displacement vector field* $\mathbf{u}(x)$ on $x(\mathcal{O})$, which is associated with the diffeomorphism $y(x)$ by way of:

$$\mathbf{u}(x) = y(x) - x. \tag{2.10}$$

One immediately sees that when $M$ has no affine structure the right-hand side of the definition is absurd. Moreover, if one looks at the "components" of $\mathbf{u}(x)$:

$$u^\mu(x) = y^\mu(x) - x^\mu, \tag{2.11}$$

one sees that when one changes to a different coordinate system around both $x$ and $y$ the new components of $\mathbf{u}$ are obtained from the coordinate *diffeomorphism* directly, rather than from its *differential*, as they should be. Hence, the "vector field" is not really a vector field in the global sense, but only a locally-defined object that does not transform properly under coordinate changes. Hence, we shall use the displacement vector field $\mathbf{u}(x)$ only with such caveats, and for the sake of consistency with the established treatments of strain.

Since, by differentiation, one has:

$$y^\mu_{,\nu} = \delta^\mu_\nu + u^\mu_{,\nu}, \tag{2.12}$$

one can then see that the components of the Cauchy-Green finite strain tensor take the form:

$$E_{\mu\nu} = u_{\mu,\nu} + u_{\nu,\mu} + g_{\kappa\lambda} u^\kappa_{,\mu} u^\lambda_{,\nu} \qquad (u_{\mu,\nu} \equiv g_{\mu\kappa} u^\kappa_{,\nu}). \tag{2.13}$$

Now suppose that the deformed state $\bar{x}(\mathcal{O})$ is the end result of a differentiable one-parameter family of embeddings $x_t(\mathcal{O})$, with $t \in [0, 1]$; i.e., $x(\mathcal{O}) = x_0(\mathcal{O})$, $\bar{x}(\mathcal{O}) = x_1(\mathcal{O})$, For instance, one could define a differentiable map $Y$: $[0, 1] \times \mathcal{O} \to M$ such that $x_t(a) = Y(t, a)$ is an embedding for every $t$; such maps are called *isotopies* in differential topology. Of course, it is now possible that $p+1 > n$, so one does not specify that $Y$ itself must be an embedding any longer. For the cases in which $p + 1$ is not greater than $n$ and $Y$ is an embedding, one sees that, in a sense, dynamics is just the statics of objects whose parameter dimensions include time. For instance, the dynamics of a point and the statics of a curve segment are both described by objects of dimension one, while the dynamics of a curve segment and the statics of a surface element are both described by objects of dimension two. One sees that what relativity theorists call a "world tube" amounts to an embedding of $[0, 1] \times \mathcal{O}$ in a Lorentzian manifold $M$.

For each value of $t \in [0, 1]$, one can pull back the metric $g$ on $x_t(\mathcal{O})$ to a metric $\bar{g}_t$ on $x(\mathcal{O})$. By the differentiability assumption, we can define the *infinitesimal strain* tensor on $x(\mathcal{O})$ to be:

$$e = \left.\frac{d\bar{g}_t}{dt}\right|_{t=0} = \lim_{t\to 0}\left(y_t^* g - g\right). \tag{2.14}$$

Its components then become:

$$e_{\mu\nu} = u_{\mu,\nu} + u_{\nu,\mu} \tag{2.15}$$

for a flat metric. One can then say that:

$$E_{\mu\nu} = e_{\mu\nu} + g_{\kappa\lambda} u^{\kappa}_{,\mu} u^{\lambda}_{,\nu} . \tag{2.16}$$

Hence, the infinitesimal strain tensor approximates the finite one up to quadratic terms in the displacement.

When $g$ is not flat, we first define the *velocity vector field* on $Y([0, 1]\times\mathcal{O})$:

$$\mathbf{v}(t, x) = \frac{\partial Y}{\partial t} . \tag{2.17}$$

We then see that the definition of the infinitesimal strain tensor can be expressed as:

$$e = \mathrm{L}_{\mathbf{v}} g, \tag{2.18}$$

in which $\mathrm{L}_{\mathbf{v}}$ refers to the Lie derivative operator associated with the vector field $\mathbf{v}$.

By Killing's theorem:

$$e_{\mu\nu} = v_{\mu;\,\nu} + v_{\nu;\,\mu} , \tag{2.19}$$

in which the semi-colon refers to the covariant derivative that is defined by the Levi-Cività connection for $g$.

When $e$ vanishes, the definition, in the present form, is also the definition of $\mathbf{v}$ as a *Killing vector field*, whose flow $y_t$ then consists of isometries of $g$. Hence, in order for there to be deformation, these diffeomorphisms cannot all preserve the metric.

One can also verify directly that when $e = 0$ the only possible analytic displacements $u^i$ are infinitesimal rigid motions of $x(\mathcal{O})$. Suppose:

$$u_\mu(x) = \sum_{|\alpha|=0}^{\infty} \frac{1}{\alpha!} a_{\mu\alpha} x^{\alpha} = a_\mu + a_{\mu\kappa} x^{\kappa} + \tfrac{1}{2}\, a_{\mu\kappa\lambda}\, x^{\kappa} x^{\lambda} + \dots \tag{2.20}$$

in which $\alpha$ is a multi-index [1].

Hence:

$$u_{\mu,\,\nu}(x) = a_{\mu\nu} + a_{\mu\nu\kappa} x^{\kappa} + \dots \tag{2.21}$$

If $e = 0$ then:

$$0 = u_{\mu,\,\nu} + u_{\nu,\,\mu} = a_{(\mu\nu)} + a_{(\mu\nu)\kappa} x^{\kappa} + \dots, \tag{2.22}$$

---

[1] Recall the basic definitions of multi-indices: the multi-index notation $\alpha$ is an abbreviation for $\alpha_1 \dots a_k$ ; its *length* $|\,\alpha\,|$ is then $k$. One defines $a! = \alpha_1 ! \dots a_k !$ and $x^{\alpha} = x^{\alpha_1} \cdots x^{\alpha_k}$ .

in which the parentheses refer to (twice) the symmetric parts of the components in the indices enclosed by them. This then implies that $a_\mu$ is arbitrary and all of the coefficients $a_{\mu\nu}$, $a_{\mu\nu\kappa\dots}$ are anti-symmetric in $\mu\nu$. However, since they are completely symmetric in all of the indices past $\mu$, one can show that they all must vanish, except for $a_{\mu\nu}$. If we suggestively re-notate the anti-symmetric matrix $a_{\mu\nu}$ by $\omega_{\mu\nu}$ to indicate an infinitesimal rotation then we have that the most general displacement that produces no infinitesimal strain takes the form:

$$u_\mu(x) = a_\mu + \omega_{\mu\nu} x^\nu; \tag{2.23}$$

i.e., an infinitesimal rigid motion.

A similar argument shows that a state of *constant* strain is define only up to infinitesimal affine transformations of the form:

$$u_\mu(x) = a_\mu + (e_{\mu\nu} + \omega_{\mu\nu})x^\nu, \tag{2.24}$$

in which the components $e_{\mu\nu}$ must represent the constant infinitesimal strain that results by symmetrized differentiation. One can then say that the most general displacement that produces a state of constant infinitesimal strain is an infinitesimal affine transformation.

## 2.2 Strain compatibility conditions

In many cases, what represents an identity from one starting point might represent an equation from another; this is generally the case with over-determined systems of equations. If one regards a system of equations – say $x^\mu = x^\mu(a^i)$ – as the result of some map $x$: $\mathbf{R}^p \to \mathbf{R}^n$ then the issue of determinacy is simply that of characterizing the image of $\mathbf{R}^p$ in $\mathbf{R}^n$ under the action of $x$. If $x$ is a surjection, so the image of $x$ is all of $\mathbf{R}^n$, then one can solve the equations for $a^i$ – but not necessarily uniquely – for any choice of $x^\mu$ in $\mathbf{R}^n$. If there are generally more than one solution for a choice of $\mathbf{R}^n$ then the system is *under-determined;* for instance, if $p > M$ then this is usually the case, depending upon the nature of $y$. If the image of $x$ is a proper subset of $\mathbf{R}^n$ then there will be $x^\mu$ in $\mathbf{R}^n$ for which no solution to the system exists; the system is then *over-determined*.

When a system is over-determined, the next task is invariably that of characterizing the nature of the image of the map that defines it. Of particular interest are the cases in which the image is itself the solution set of some further system of equations – either algebraic or differential – such as the kernel of a linear map.

One elementary example is given by the exterior differential equation $d\alpha = \beta$, where $\alpha$ is a $k$-form and $\beta$ is a $k$+1-form. Since the image of the linear differential operator $d$: $\Lambda^k M \to \Lambda^{k+1} M$ does not consist of all $k$+1-forms, there will be $k$+1-forms $\beta$ such that the equation has no solution. However, because $d^2 = 0$, in any case, one sees that any solution to the given equation will also be a solution to the equation $d\beta = 0$; i.e., it will be in the kernel of $d$: $\Lambda^{k+1} M \to \Lambda^{k+2} M$. According to the Poincaré lemma, there will be a neighborhood of any point of $M$ for which the converse statement is true for the local representative of $\beta$; viz., if it is a solution to the exterior differential equation $d\beta = 0$ then there will be a solution to the exterior differential equation $d\alpha = \beta$. However, de Rham's theorem says that this converse is true only if the real cohomology of $M$ vanishes in dimension $k$+1. In particular, when $k = 0$ the topological issue is simple connectivity.

In the cases where the map $x$ takes the form of a differential operator, so the system of equations is a system of ordinary differential equations, the system of equations that define its image subspace are referred to as either *integrability conditions* or *compatibility conditions.* Although continuum mechanics generally prefers to use the word "compatibility" we shall sometimes use the word "integrability" to emphasize the fact that one is trying to integrate systems of differential equations.

Both the definition (2.9) of the finite strain tensor $E$ and the definition (2.15) the infinitesimal strain tensor $e$ for a given deformation $y(x)$ define over-determined systems of partial differential equations for the deformation when one starts with $E$ or $e$ as the givens. Hence, in order to say whether a solution exists for a given $E$ or $e$ one must specify a system of equations that it must solve, which then represent integrability conditions for the strain tensor, in either form.

The traditional derivation of the infinitesimal strain compatibility conditions deals with the displacement vector field, or rather, its covector field $u_i(x)$, so one attempts to construct that displacement vector field by integrating something that involves the infinitesimal strain tensor.

One starts with:

$$u_i(x) = u_{0i} + \int_{x_0}^{x} u_{ij}(\xi) d\xi^j \, , \tag{2.25}$$

in which $u_{ij}(x)$ does not have to be integrable into $u_{i,j}(x)$ for some $u_i(x)$, at this point.

We first make the replacement:

$$u_{ij} = \tfrac{1}{2} e_{ij} + \tfrac{1}{2} \omega_{ij} \qquad\qquad (\omega_{ij} = u_{ij} - u_{ji}) \tag{2.26}$$

in the previous integral to obtain:

$$u_i(x) = u_{0i} + \tfrac{1}{2} \int_{x_0}^{x} [e_{ij}(\xi) + \omega_{ij}(\xi)] d\xi^j \, . \tag{2.27}$$

In Cesaro's derivation of the infinitesimal strain compatibility conditions (cf., Volterra [**4**]), he applies integration by parts to the infinitesimal rotation part of the integral to obtain:

$$u_i(x) = u_{0i} + \tfrac{1}{2} \int_{x_0}^{x} e_{ij}(\xi) d\xi^j + \tfrac{1}{2} [\omega_{ij}(\xi)\xi_j]_{x_0}^{x} - \tfrac{1}{2} \int_{x_0}^{x} d\omega_{ij}(\xi) \xi^j \, . \tag{2.28}$$

Now, one see that if $u_{ij,k}$ is symmetric in $jk$ then its complete anti-symmetrization $u_{[ij,k]}$ must be zero. This leads to:

$$0 = \omega_{ij,k} - e_{ij,k} + e_{jk,i} \tag{2.29}$$

so:

$$d\omega_{ij} = (e_{ij,k} - e_{jk,i}) \, dx^k \tag{2.30}$$

which makes:

$$du_{ij} = \tfrac{1}{2}(de_{ij} + d\omega_{ij}) = \tfrac{1}{2}(e_{ij,k} + e_{jk,i} - e_{ki,j})\, dx^k , \tag{2.31}$$

and the integral (2.28) takes the form:

$$u_i(x) = u_{0i} + \tfrac{1}{2}[\omega_{ij}(\xi)\xi^j]_{x_0}^{x} + \tfrac{1}{2}\int_{x_0}^{x}[e_{ij} - (e_{ij.k} - e_{jk.i})\xi^k]d\xi^j , \tag{2.32}$$

which we rewrite as:

$$[u_i(\xi) - \tfrac{1}{2}\omega_{ij}(\xi)\xi^j]_{x_0}^{x} = \tfrac{1}{2}\int_{x_0}^{x}[e_{ij} - (e_{ij,k} - e_{jk,i})\xi^k]d\xi^j , \tag{2.33}$$

Of course, in order to carry out the integration one must specify a path from $x_0$ to $x$, which then suggests the question of whether the integral is truly path-independent. This is a matter of either homotopy or de Rham cohomology. As long as one uses exact 1-forms as integrands in the right-hand side of (2.33), this is guaranteed, although it is sufficient that they be closed in order make the integral independent of the path, at least within its homotopy class. Requiring that the integrand must be a closed 1-form for each $i$ then gives the resulting integrability conditions on $e_{ij}$ , which then become the celebrated conditions of Barré de Saint Venant:

$$0 = e_{ij,k,l} - e_{jk,l,i} + e_{kl,i,j} - e_{li,j,k} . \tag{2.34}$$

Nowadays, it is customary to derive these conditions as expressing the fact that the curvature of the deformed metric that results on the final state has vanishing curvature, but we now see that one can also derive it from purely topological considerations. We shall return to this point later in this article, but first we shall present the foregoing construction in another way.

### 2. 3 Alternative form of the derivation of compatibility conditions

There are some aspects of the preceding derivation of the strain compatibility conditions that seem somewhat weak by modern standards:

1. Although one is dealing with finite loops in a finite (singular) deformation of the initial state one is nevertheless obtaining infinitesimal transformations as a consequence.
2. One might wish to avoid the use of a displacement vector field to represent a diffeomorphism or immersion.
3. The result suggests that one cannot apply other deformations besides rigid motions to the end face before identifying points when, intuitively, one could at least perform shears and dilatations, as well.

In this section, we show that all of these weaknesses can be circumvented in a straightforward way. It is closer in spirit to the approach taken by Pommaret [**12**] to recasting the Cosserat [**13**] theory of deformable bodies in the context of the Spencer sequence for the integrability of over-determined systems of partial differential equations, although we shall not explicitly introduce that formalism here.

Let us start with the initial state $x(\mathcal{O})$ of the $n$-dimensional object $\mathcal{O}$ in the $n$-dimensional $M$, and assume that there is a global coframe field:

$$\theta^i = y^i_j dx^j , \qquad (i = 1, \ldots, n). \tag{2.35}$$

defined on $x(\mathcal{O})$. Of course, this assumes that such a thing is topologically possible; i.e., that $x(\mathcal{O})$ is *parallelizable* as a manifold. Although this possibility is usually hard to come by in the general case, nonetheless, as Stiefel showed in 1936, it is *always* true for compact, orientable three-dimensional manifolds, which seem to describe the majority of material objects that one considers in continuum mechanics, unless one approximates them as membranes and strings.

Rather than introduce the displacement vector field, we then attempt to construct the diffeomorphism $y^i(x)$ directly by integration in a manner that is analogous to the previous one:

$$y^i(x) = y^i_0 + \int_{x_0}^{x} y^i_j(\xi) d\xi^j . \tag{2.36}$$

In order for the integral in (2.36) to be path-independent, one must impose the condition that each $\theta^i$ be closed, which is equivalent to the symmetry condition:

$$y^i_{j.k} = y^i_{k,j} , \tag{2.37}$$

which is automatic when $y^i_j(x)$ is integrable since the $\theta^i = dy^i$. Hence, one expects that if one can solve for $y^i_j$ in terms of $E_{ij}$ or $e_{ij}$ then the latter condition would define integrability conditions for either of the strain tensors.

The integrability problem now takes the form of asking whether one can construct a diffeomorphism $y$: $x(\mathcal{O}) \rightarrow \bar{x}(\mathcal{O})$ by starting with the local coframe field $\theta^i = y^i_j dx^j$ on $x(\mathcal{O})$ and integrating it along a curve $l(s)$ from some chosen initial point $x_0$ to all of the other points $x \in x(\mathcal{O})$ to obtain the $y^i(x)$ directly:

$$y^i(x) = y^i_0 + \int_{x_0}^{x} \theta^i \Big|_{\xi} . \tag{2.38}$$

One sees that since it is obvious that when the $\theta^i$ are not all exact no such diffeomorphism can exists, what one will actually construct by integration is a "singular diffeomorphism" or *immersion*, which can have self-intersections. That is, when one closes the curve $l(s)$ into a loop through $x_0$ there might be a jump discontinuity in the value of $y$ at $x_0$. If one looks at all of the other loops that $l$ is homotopic to then the set of all images of $x_0$ defines a "cut;" i.e., a hypersurface in $x(\mathcal{O})$ that intersects all of the loops in $[l]$ transversally.

There are two ways that each $\theta^i$ might not be exact: It could be closed, but not exact, or it could be not closed. It is the former possibility that is of interest to homotopy, since when the $\theta^i$ are all closed, but not exact, the value of the integral (2.38) in independent of homotopy. Hence, the value of $[y^i]$ is unambiguously associated with either the homotopy class $[l]$ or the de Rham cohomology classes of the $\theta^i$.

Let us represent the $\theta^i$ in a different manner. From the product rule for exterior differentiation, we have:

$$\theta^i = d(y^i_j x^j) - dy^i_j x^j = d\psi^i + \gamma^i_j \psi^j \equiv \nabla \psi^i, \tag{2.39}$$

in which we have introduced the "quasi-coordinate" $\psi^i = y^i_j x^j$, the linear "connection 1-form":

$$\gamma^i_j = -\, dy^i_k \tilde{y}^k_j\,, \tag{2.40}$$

and the "exterior covariant differential" $\nabla$; we shall, in due course, justify the use of the last two terms in quotes.

We then have:

$$\left[y^i - y^i_j \xi^j\right]^x_{x_0} = \int_{x_0}^x \gamma^i_j \psi^j\ . \tag{2.41}$$

If the integral is homotopy-invariant then the integrand must also be a closed 1-form:

$$0 = d(\gamma^i_j \psi^j) = -\, d\theta^i; \tag{2.42}$$

that is, the coframe members $\theta^i$ must all be closed. If they were exact and of the form $\theta^i = dy^i$ then one would obtain the desired diffeomorphism $y$ by quadrature. However, if $x(\mathcal{O})$ is multiply connected then there will be closed 1-forms that are not exact.

Now, let us close the curve $l(s)$ into a loop $l_0$ by making $x = x_0$ :

$$\left[y^i - y^i_j x^j_0\right] = \int_{l_0} \gamma^i_j \psi^j\ . \tag{2.43}$$

If $x(\mathcal{O})$ is simply connected then the integral must vanish, along with the jump discontinuity $\left[y^i - y^i_j x^j_0\right]$ at $x_0$ associated with the loop $l_0$ . However, if $x(\mathcal{O})$ is multiply connected then the integral, as well as the jump discontinuity, can be non-vanishing. In particular, there will be as many non-homotopic loops at $x_0$ as the cardinality of the fundamental group $\pi_1(x(\mathcal{O}), l_0)$.

Since:

$$\gamma_{ij}\ \psi^j = dy_{ij}\ x^j \tag{2.44}$$

and:

$$dy_{ij} = d(\delta_{ij} + u_{ij}) = du_{ij}\,, \tag{2.45}$$

from (2.31), we can rewrite (2.41) in the form:

$$[y_i(\xi) - y_{ij}(\xi)\xi^j]_{x_0}^x = +\ \tfrac{1}{2}\int_{x_0}^x [e_{ij,k} + e_{jk,i} - e_{ki,j}]\xi^j d\xi^k\,, \tag{2.46}$$

and the integrability conditions also take the form of demanding that the 1-forms:

$$\varepsilon_i = \tfrac{1}{2}(e_{ij,k} + e_{jk,\,i} - e_{ki,j})\ \xi^j\ d\xi^k, \tag{2.47}$$

must be closed.

By direct computation, one sees that this also gives the Saint Venant conditions that are expressed in (2.34).

Hence, we have arrived at the same strain compatibility conditions by addressing the deformation $y^i(x)$ directly, rather than the displacement vector field $\mathbf{u}(x)$ that represents it and have obtained finite transformations for the jump discontinuity that comes about when integrating around a homotopically non-trivial loop, rather than infinitesimal ones. However, we also see that the transformation in question can now include all affine transformations, and not merely all rigid motions.

## 3 The nature of the connection $\gamma^i_j$

So far, we have not explicitly introduced a connection that would allow us to speak of holonomy and parallel translation. However, we have introduced one implicitly in the form of the matrix-valued 1-form $\gamma^i_j$, which is associated with the set of $n$ linearly independent 1-forms $\theta^i$, which defined in (2.35), and which then constitute a coframe field on $x(\mathcal{O})$. Hence, let us recall some of the geometric constructions that are intrinsic to frame fields and coframe fields.

### 3.1 Structure functions of frame fields

Let $\{\mathbf{e}_i(x), i = 1, \ldots, n\}$ be a an $n$-frame field on an open subset $U \subset x(\mathcal{O})$; i.e., a set of $n$ linearly independent vector fields on $U$. Since they collectively span each tangent space on $U$, any tangent vector can be expressed as a linear combination of the frame members, and, in particular, so can every Lie bracket $[\mathbf{e}_i, \mathbf{e}_j]$. There will then be a set of $n^3$ *structure functions* $c^k_{ij}(x)$ such that:

$$[\mathbf{e}_i, \mathbf{e}_j] = c^k_{ij}\mathbf{e}_k\,. \tag{3.1}$$

The anti-symmetry and Jacobi identity for the Lie bracket imply the following symmetries of the structure functions:

$$c_{ij}^k = -c_{ji}^k, \qquad c_{ij}^l c_{lk}^m + c_{jk}^l c_{li}^m + c_{ki}^l c_{lj}^m = 0. \tag{3.2}$$

The simplest case of $c_{ij}^k$ is when they all vanish. One calls such a local frame field *holonomic*, and, in the contrary case, *anholonomic*. The $n$-dimensional Lie algebra that is defined by such a local frame field is simply the Abelian Lie algebra on $\mathbf{R}^n$ that represents the infinitesimal generators of translations.

The fundamental example of a holonomic frame field is the *natural* frame field $\{\partial/\partial x^i, i = 1, \ldots, n\}$ that is defined by any coordinate chart $(U, x^i)$ on the manifold $M$. A first question to ask is whether every holonomic frame field $\mathbf{e}_i$ on $U$ can be integrated into a natural frame field, which amounts to the construction of the diffeomorphism $x^i: U \to \mathbf{R}^n$, $x \mapsto x^i(x)$ by starting with the local frame field $\mathbf{e}_i$.

A first obstruction to this is clearly when the open subset $U$ is not diffeomorphic to $\mathbf{R}^n$. Of particular interest to us is the case in which $U$ is multiply connected.

One finds that the concerns of topology become more self-evident when one switches to the reciprocal coframe field $\theta^i$ to $\mathbf{e}_i$, which is the set of $n$ linearly independent covector fields that are uniquely defined by the requirement:

$$\theta^i(\mathbf{e}_j) = \delta^i_j. \tag{3.3}$$

The reciprocal coframe field to a natural frame field $\partial/\partial x^i$ consists of the 1-forms $dx^i$.

If one applies the intrinsic formula for the exterior derivative [**14, 15**] of a 1-form $\alpha$:

$$d\alpha(X, Y) = \alpha(X)Y - \alpha(Y)X - \alpha[X, Y] \tag{3.4}$$

to the $\theta^i$ then one obtains:

$$d\theta^k(\mathbf{e}_i, \mathbf{e}_j) = \mathbf{e}_i(\theta^k(\mathbf{e}_j)) - \mathbf{e}_j(\theta^k(\mathbf{e}_i)) - \theta^k[\mathbf{e}_i, \mathbf{e}_j] = -\theta^k[\mathbf{e}_i, \mathbf{e}_j] = -c_{ij}^k, \tag{3.5}$$

or:

$$d\theta^k = -\tfrac{1}{2} c_{ij}^k\, \theta^i \wedge \theta^j, \tag{3.6}$$

which generalize the Maurer-Cartan equations for a right-invariant coframe field on a Lie group. In that event, the structure functions are constant, due to right-invariance.

Hence, we see that the local frame field $\mathbf{e}_i$ is holonomic iff all of the 1-forms $\theta^i$ in its reciprocal coframe field are closed. We also see that the 1-forms of a natural frame field $dx^i$ are exact. Hence, if a holonomic frame field is defined on an open subset $U$ in $M$ that is not simply connected then its reciprocal coframe field might very well be closed, but not exact, in which case it represents de Rham cohomology class in dimension one. This is, in fact, the situation that we are confronted with regarding the $\theta^i$.

The simplest way to obtain anholonomic frame fields is to start with a holonomic frame field $\partial/\partial x^i$ and subject it to the action of a smooth function $y$: $U \to GL(n)$, $x \mapsto y^i_j(x)$:

$$\mathbf{e}_i(x) = \tilde{y}^j_i(x)\frac{\partial}{\partial x^j}; \tag{3.7}$$

the reason that we use the inverse matrix is because we shall be more concerned with the reciprocal coframe field $\theta^i = y^i_j dx^j$.

One then has:

$$d\theta^i = dy^i_j \wedge dx^j = dy^i_k \tilde{y}^k_j \wedge \theta^j = -\gamma^i_j \wedge \theta^j. \tag{3.8}$$

### 3.2 Teleparallelism connection in anholonomic frame fields

Note that, if $d$ refers to the ordinary differential this time, and not the exterior derivative, then:

$$d\theta^i = dy^i_j \otimes dx^j = -\gamma^i_j \otimes \theta^j. \tag{3.9}$$

This provokes us to define a covariant differential operator on the frame field and its reciprocal coframe field in the form:

$$\nabla\mathbf{e}_i = d\mathbf{e}_i - \gamma^j_i \otimes \mathbf{e}_j, \qquad \nabla\theta^i = d\theta^i + \gamma^i_j \otimes \theta^j, \tag{3.10}$$

which then vanish for this definition of $\gamma^i_j$.

Hence, if we wish to interpret the $\mathfrak{gl}(n)$-valued 1-form $\gamma^i_j$ as the local representative of a linear connection on $U$ then we can say that the 1-forms $\gamma^i_j$ represent the linear connection that makes the local frame field $\mathbf{e}_i$ parallel, as well as its reciprocal coframe field $\theta^i$. To relativity physicists since Einstein, this connection has come to be called the *teleparallelism connection.* It is intriguing that Einstein and Mayer gave up on teleparallelism as a way of unifying gravitation and electromagnetism into a single field theory several years before Stiefel published the first definitive treatment of the topological obstructions to the existence of teleparallism.

We extend the operator $\nabla$ to vector fields and covector fields by saying that if $\mathbf{v} = \bar{v}^i\mathbf{e}_i$ and $\alpha = \bar{\alpha}_i\theta^i$ then:

$$\nabla\mathbf{v} = d\bar{v}^i \otimes \mathbf{e}_i + \bar{v}^i\nabla\mathbf{e}_i = d\bar{v}^i \otimes \mathbf{e}_i, \tag{3.11}$$

$$\nabla\alpha = d\bar{\alpha}_i \otimes \theta^i + \bar{\alpha}_i\nabla\theta^i = d\bar{\alpha}_i \otimes \theta^i. \tag{3.12}$$

This means that a vector field or covector field is parallel with respect to this connection iff its components with respect to the frame field or its reciprocal coframe, resp., are constants. In particular, the tangent vectors to a geodesic appear to lie in a "straight line."

From the Cartan structure equations [**5, 15**], the torsion and curvature 2-forms associated with $\gamma^i_j$ are [1]:

$$\bar{\Theta}^i = d_\wedge \theta^i + \gamma^i_j \wedge \theta^j = dy^i_j \wedge dx^j - dy^i_k \tilde{y}^k_l \wedge y^l_j dx^j = 0, \tag{3.13}$$

$$\bar{\Omega}^i_j = d_\wedge \gamma^i_j + \gamma^i_k \wedge \gamma^k_j = dy^i_k \wedge d\tilde{y}^k_j + dy^i_k \tilde{y}^k_l \wedge dy^l_m \tilde{y}^m_j = 0. \tag{3.14}$$

The latter result follows from the fact that $y^i_k \tilde{y}^k_j = \delta^i_j$, which then implies that:

$$d\tilde{y}^i_j = -\tilde{y}^i_k dy^k_l \tilde{y}^l_j. \tag{3.15}$$

The Bianchi identities take the general form:

$$\nabla_\wedge \Theta^i = d_\wedge \Theta^i + \gamma^i_j \wedge \Theta^j = \Omega^i_j \wedge \theta^j, \qquad \nabla_\wedge \Omega^i_j = d_\wedge \Omega^i_j + \gamma^i_k \wedge \Omega^k_j = 0, \tag{3.16}$$

so for the present connection, they are then trivial.

One finds, in fact, that the Bianchi identity for the curvature is the dual of the Jacobi identity for the commutation relations (3.1) when the local frame field $\mathbf{e}_i$ is reciprocal to the coframe field $\theta^i$, just as the equations for the structure functions in terms of the coframe field are dual to the commutation relations of the frame field.

Since we have introduced two different metrics $g$ and $\bar{g}$, we should examine the non-metricity tensors that are associated with each.

One can characterize the deformed metric $\bar{g}$ by the fact that it makes the anholonomic coframe field $\theta^i$ orthonormal:

$$\bar{g} = \delta_{ij}\, \theta^i\, \theta^j. \tag{3.17}$$

The non-metricity tensor defined by $\bar{g}$ then clearly vanishes

$$\bar{Q} \equiv \nabla \bar{g} = \delta_{ij}\, (\nabla\theta^i\, \theta^j + \theta^i \nabla\theta^j) = 0 \quad . \tag{3.18}$$

Since we have already established that the torsion $\bar{\Theta}^i$ of $\nabla$ vanishes for the anholonomic frame, one sees that this implies, by uniqueness, that $\gamma^i_j$ must represent the Levi-Cività connection for the deformed metric.

Now, let us we represent the ambient metric $g$ in the anholonomic coframe field as:

$$g = g_{ij}\, \theta^i\, \theta^j \qquad (g_{ij} = \delta_{kl} \tilde{y}^k_i \tilde{y}^l_j) \tag{3.19}$$

---

[1] From now on, we denote the exterior derivative operator by $d_\wedge$ to avoid confusing it with the differential map $d$.

If we do a polar decomposition of the matrix $\tilde{y}^i_j$ into a product:

$$\tilde{y}^i_j = \tilde{E}^i_k \tilde{R}^k_j \tag{3.20}$$

of a shear $\tilde{E}^i_j$ and a rotation $\tilde{R}^i_j$ then:

$$g_{ij} = \delta_{kl} \tilde{y}^k_m \tilde{y}^l_j = \delta_{kl} \tilde{E}^k_m \tilde{E}^l_j , \tag{3.21}$$

since rotation matrices will have the property:

$$\delta_{kl} \tilde{R}^k_m \tilde{R}^l_j = \delta_{ij} . \tag{3.22}$$

Thus, the part of $\tilde{y}^i_j$ that is responsible for the change in the metric is the shear.

The non-metricity tensor that $g$ defines is then:

$$Q \equiv \nabla g = dg_{ij} \otimes \theta^i \theta^j, \tag{3.23}$$

which does not have to vanish, depending upon the nature of $\tilde{y}^i_j$. In fact:

$$\begin{aligned} dg_{ij} &= \delta_{kl} (d\tilde{y}^k_i \tilde{y}^l_j + \tilde{y}^k_i d\tilde{y}^l_j) \\ &= -\delta_{kl} (\tilde{y}^k_m dy^m_n \tilde{y}^n_i \tilde{y}^l_j + \tilde{y}^k_i \tilde{y}^l_m dy^m_n \tilde{y}^n_j) \\ &= (\delta_{kl} \tilde{y}^k_m \tilde{y}^l_j) \gamma^m_i + (\delta_{kl} \tilde{y}^k_i \tilde{y}^l_m) \gamma^m_j , \end{aligned}$$

which makes:

$$dg_{ij} = g_{ik} \gamma^k_j + g_{jk} \gamma^k_i = \gamma_{ij} + \gamma_{ji} . \tag{3.24}$$

Hence, non-vanishing non-metricity is due to the non-vanishing of the symmetric part of $\gamma_{ij}$, which is, in turn, the infinitesimal generator of the shear $\tilde{E}^i_j$.

### 3.2 Teleparallelism connection relative to the holonomic frame field

We see that the connection $\gamma^i_j$ makes geometry appear Euclidian in the anholonomic frame geometry. Now, let us look at how the same geometry appears in the holonomic frame field $\partial/\partial x^i = y^j_i \mathbf{e}_j$ and its reciprocal coframe field $dx^i = \tilde{y}^i_j \theta^j$.

First, we see that:

$$\nabla(\partial/\partial x^i) = d(\partial/\partial x^i) - \gamma^j_i \otimes (\partial/\partial x^j) = dy^j_k \tilde{y}^k_i \otimes (\partial/\partial x^j) = y^k_{l,j} \tilde{y}^l_i \, dx^j \otimes (\partial/\partial x^k), \tag{3.25}$$

$$\nabla(dx^i) = d(dx^i) + \gamma^i_j \otimes dx^j = - dy^i_k \tilde{y}^k_j \otimes dx^j = - y^i_{l,j} \tilde{y}^l_k \, dx^j \otimes dx^k. \tag{3.26}$$

The components of the connection 1-form $\gamma^i_j$ in this frame field are then:

$$\gamma^i_{jk} = -y^i_{l,j}\tilde{y}^l_k . \tag{3.27}$$

The covariant differentials of a vector field $\mathbf{v} = v^i\, \partial/\partial x^i$ and a covector field $\alpha = \alpha_i\, dx^i$ are then:

$$\nabla\mathbf{v} = dv^i \otimes \partial/\partial x^i + v^i\nabla(\partial/\partial x^i) = (dv^i + \gamma^i_j v^j) \otimes \partial/\partial x^i, \tag{3.28}$$

$$\nabla\alpha = d\alpha_i \otimes dx^i + \alpha_i\, \nabla(dx^i) = (d\alpha_i - \gamma^j_i\, \alpha_j) \otimes dx^i. \tag{3.29}$$

One can compare these expressions with (3.11) and (3.12), and note that they imply that $\nabla$ does indeed represent a "covariant" derivative in the holonomic frame field.

Thus, a vector field or covector field that appears parallel in the anholonomic frame $\mathbf{e}_i$ will have component differentials:

$$dv^i = -\,\gamma^i_j v^j, \qquad\qquad d\alpha_i = \gamma^j_i\, \alpha_j, \tag{3.30}$$

in the holonomic frame, which are not always constant. In fact, the only way that the frame transition function $y^i_j$ will take constant components to other constant components is if it is also constant.

The torsion and curvature 2-forms for $\gamma^j_i$, relative to the holonomic frame are:

$$\Theta^i = d_\wedge(dx^i) + \gamma^j_i \wedge dx^j = -\, dy^i_k\tilde{y}^k_j \wedge dx^j = -\, dy^i_k\tilde{y}^k_j \wedge \tilde{y}^j_l\theta^l = \tilde{y}^j_j (d_\wedge\theta^j) \tag{3.31}$$

$$\Omega^i_j = d\gamma^i_j + \gamma^i_k \wedge \gamma^k_j = 0. \tag{3.32}$$

Hence, the curvature still vanishes, because the connection has not changed intrinsically, but since we are considering a different local coframe field, the torsion is different. It has components:

$$\Theta^i_{jk} = y^i_{l,j}\tilde{y}^l_k - y^i_{l,k}\tilde{y}^l_j . \tag{3.33}$$

in the holonomic coframe field.

In fact, the torsion components $\Theta^i_{jk}$ in this coframe field are related to the structure functions $c^i_{jk}$ of $\theta^i$ by the fact that:

$$d_\wedge\theta^i = -\,\tfrac{1}{2}c^i_{jk}\,\theta^j \wedge \theta^k = -\tfrac{1}{2}(c^i_{lm}y^l_j y^m_k)dx^j \wedge dx^k . \tag{3.34}$$

Hence:

$$\Theta^i_{jk} = -\, y^i_l y^m_j y^n_k c^l_{mn} , \tag{3.35}$$

which is just the tensorial transformation law for the components.

Thus, in this holonomic frame field, the geometry no longer appears flat, in general.

Now, let us examine the non-metricity tensors that are defined by $g$ and $\bar{g}$, with:

$$g = \delta_{ij}\, dx^i\, dx^j, \qquad \bar{g} = \bar{g}_{ij} dx^i dx^j \qquad (\bar{g}_{ij} = \delta_{kl} y_i^k y_j^l), \tag{3.36}$$

this time.

The non-metricity tensor defined by $g$ is now [1]:

$$\begin{aligned} Q \equiv \nabla g &= \delta_{ij}\,(\nabla(dx^i) \odot dx^j + dx^i \odot \nabla(dx^j)) \\ &= -\delta_{ij}(\gamma_k^i \otimes dx^k \odot dx^j + dx^i \odot \gamma_k^j \otimes dx^k), \\ &= -(\delta_{ik}\gamma_j^k + \delta_{jk}\gamma_i^k) \otimes dx^i \odot dx^j, \end{aligned}$$

or:

$$Q = -(\gamma_{ij} + \gamma_{ji}) \otimes dx^i \odot dx^j. \tag{3.37}$$

Thus, the non-metricity tensor of $g$ is still carried by the symmetric part of $\gamma_{ij}$, except that this time we have lowered the index with the components of $g$ in the holonomic frame.

The non-metricity tensor defined by $\bar{g}$ now takes the form:

$$\bar{Q} \equiv \nabla \bar{g} = (d\bar{g}_{ij} - \bar{g}_{ik}\gamma_j^k - \bar{g}_{jk}\gamma_i^k) \otimes dx^i \odot dx^j, \tag{3.38}$$

but since it is, after all, a tensor field, its vanishing in one frame field should imply its vanishing in any other.

Hence $\bar{Q} = 0$, or:

$$d\bar{g}_{ij} = \bar{g}_{ik}\gamma_j^k + \bar{g}_{jk}\gamma_i^k. \tag{3.39}$$

## 4 The connection used in the Saint Venant condition

The expression $\frac{1}{2}(e_{ij,k} + e_{jk,i} - e_{ki,j})$ that appears in (2.31) bears an uncanny resemblance to the components of the Levi-Cività connection for a metric. Of course, as we pointed out above, we have also defined a Levi-Cività connection for the deformed metric $\bar{g}$ by way of the teleparallelism connection. Hence, we should expect that we are simply dealing with two different ways of obtaining the same thing.

In fact, if one lets the ambient metric $g$ in $M$ be Euclidian then:

$$g = \delta_{ij}\, dx^i\, dx^j \tag{4.1}$$

and the deformed metric $\bar{g}$ has holonomic components:

$$\bar{g}_{ij} = \delta_{ij} + E_{ij}. \tag{4.2}$$

---

[1] We now insert the symbol $\odot$ to represent the symmetrized tensor product, for clarity.

The Levi-Cività connection then has the holonomic components:

$$\bar{\Gamma}^i_{jk} = \tfrac{1}{2}\bar{g}^{il}(g_{lj,k} + g_{jl,k} - g_{jk,l}) = \tfrac{1}{2}\bar{g}^{il}(g_{lj,k} + g_{jl,k} - g_{jk,l}) . \tag{4.3}$$

If one expresses the inverse metric to $\bar{g}$ in the form:

$$\bar{g}^{ij} = \delta^{ij} + E^{ij} \tag{4.4}$$

then one must have, as a consequence:

$$\delta^i_j = \bar{g}^{ik}\bar{g}_{kj} = \delta^i_j + E^{ik}\delta_{kj} + \delta^{ik}E_{kj} + E^{ik}E_{kj} ; \tag{4.5}$$

i.e.:

$$E^{ij} = - \delta^{ik}\, \delta^{jl}\, E_{kl} + \delta^{jl}\, E^{ik}\, E_{kl} . \tag{4.6}$$

If one approximates $E_{ij}$ by $e_{ij}$ and $E^{ij}$ by $e^{ij}$, and ignores the quadratic terms, then this takes the form:

$$e^{ij} = - \delta^{ik}\, \delta^{jl}\, e_{kl} , \tag{4.7}$$

and the components $\bar{\Gamma}^i_{jk}$ , up to quadratic terms, take the approximate form:

$$\bar{\Gamma}^i_{jk} \approx \tfrac{1}{2}\delta^{il}\,(e_{lj,\,k} + e_{jk,\,l} - e_{kl,\,j}), \tag{4.8}$$

which is essentially the expression in (2.31) with an index raised by the ambient metric, rather than the deformed one.

If we define the connection 1-form:

$$\bar{\Gamma}^i_j = \bar{\Gamma}^i_{jk}\, dx^k = \tfrac{1}{2}\delta^{il}\,(e_{lj,\,k} + e_{jk,\,l} - e_{kl,\,j})\, dx^k \tag{4.9}$$

then the Riemanian curvature 2-form:

$$\bar{\Omega}^i_j = \tfrac{1}{2}R^i_{jkl}dx^k \wedge dx^l \tag{4.10}$$

equals:

$$\bar{\Omega}^i_j = d\bar{\Gamma}^i_j + \bar{\Gamma}^i_k \wedge \bar{\Gamma}^k_j . \tag{4.11}$$

If we lower the upper index with $\delta_{ij}$ then we find that:

$$d\bar{\Gamma}_{ij} = -\tfrac{1}{4}(e_{jk,\,i,\,l} - e_{jl,\,i,\,k} - e_{ki,\,j,\,l} + e_{li,\,j,\,k})\, dx^k \wedge dx^l, \tag{4.12}$$

since:

$$e_{ij,\,[k,\,l]} = 0. \tag{4.13}$$

As the components of $d\bar{\Gamma}_{ij}$ are essentially the right-hand side of (2.34), we see that, once again, up to quadratic terms, the vanishing of $\bar{\Omega}^i_j$ is equivalent to the Saint Venant necessary conditions for infinitesimal strain compatibility.

In either event, holonomic or anholonomic, the deformed connection is still flat; i.e., both its torsion and curvature vanish.

If we return to the condition that we imposed on $u_{ij,k}$ in order to obtain the relationship between $d\omega_{ij}$ and $de_{ij}$ – namely, that it was symmetric in $j,k$ – then we see that this is really the condition that the 1-forms $\upsilon_i = u_{ij}\, dx^j$ all be closed. However, since:

$$du_{ij} = dy_{ij} \tag{4.14}$$

we can also see that this is equivalent to the condition that the coframe members $\theta^i$ all be closed, which we understand to imply that the integration in question is homotopy-invariant.

The fact that the complete anti-symmetrization of $u_{ij,k}$ then vanishes then amounts to the vanishing of the 3-form:

$$U = d\upsilon_i \wedge dx^i = -\, dy_{[ij]} \wedge dx^i \wedge dx^j\,. \tag{4.15}$$

However, this can vanish even when $d\upsilon_i$ is non-vanishing.

If we set:

$$d\upsilon_i = d\omega_i + de_i \tag{4.16}$$

then the vanishing of $U$ gives:

$$d\omega_i \wedge dx^i = -\, de_i \wedge dx^i, \tag{4.17}$$

which then becomes the relation (2.30).

The essential point of the foregoing section was that this relation follows as a consequence of the assumption that all of the $\theta^i$ are closed.

## 5 Holonomy of flat connections on multiply-connected manifolds

If we return to our original problem of the integrability of $\theta^i$ into a diffeomorphism then we sees that the jump discontinuity at $x_0$:

$$\left[y^i - y^i_j x^j_0\right] = \int_{l_0} \gamma^i_j \psi^j \tag{5.1}$$

originates in two contributions:

$$[y^i] = \int_{l_0} \theta^i\,, \qquad \left[y^i_j x^j_0\right] = \left[y^i_j\right] x^j_0 = -\int_{l_0} \gamma^i_j \psi^j\,. \tag{5.2}$$

If we think of the ordered pair $(\theta^i, \gamma^i_j) \in \mathfrak{a}(n)$, the Lie algebra of the affine group $A(n)$, as defining an affine connection on $x(\mathcal{O})$ then, at first sight, the relations (5.2) seem to represent the holonomy element that $(y^i, -y^i_j) \in A(n)$ that one associates with the loop $l_0$ by means of parallel translation using the connection $\gamma^i_j$. However, one must recall that we said that $\nabla\psi^j$ equals $\theta^i$, not zero. Hence, we are not actually parallel-translating the vector $\psi^j\mathbf{e}_i$ by means of the connection $\gamma^i_j$, so $h(l_0)$ is not equal to $(y^i, -y^i_j)$. Nevertheless, since the literature of dislocations and disclinations is phrased in the language of holonomy, let us examine the holonomy group element $h(l_0) = (A^i, B^i_j) \in A(n)$ that *is* associated with the loop $l_0$ by way of the connection $\gamma^i_j$.

Since the coframe field $\theta^i$ is assumed to be parallel for the connection $\gamma^i_j$, the translational part $A^i$ of the holonomy group element $h(l_0) = (A^i, B^i_j) \in A(n)$ is still given by the first expression in (5.2). However, the linear part $B^i_j$ is obtained as follows:

If one parallel translates a linear frame $\mathbf{e}_i(s)$, $s \in [0, 1]$ around $l_0$ ($l(1) = l(0)$) using $\gamma^j_i$ then $\mathbf{e}_i(s)$ must satisfy the linear system of ordinary differential equations:

$$d\mathbf{e}_i(s) = \gamma^j_i \otimes \mathbf{e}_j . \tag{5.3}$$

If one solves the equations in the form:

$$\mathbf{e}_i(s) = \mathbf{e}_j(0)\ \exp\int_0^s \gamma^i_j \tag{5.4}$$

then one can then define $h(l_0)$ by the matrix $[h(l_x)]^i_j$ that makes:

$$\mathbf{e}_i(1) = \mathbf{e}_j(0)\ [h(l_0)]^j_i . \tag{5.5}$$

Hence, $h(l_0)$ can be represented by the integral:

$$[h(l_0)]^j_i = \exp\int_{l_0} \gamma^i_j . \tag{5.6}$$

If $\gamma^i_j$ is a linear connection on $M$ then the *holonomy map* $h$: $\Omega M_x \to GL(n)$ at each $x \in M$ takes each loop $l_x \in \Omega_x M$ through $x$ to the element $h(l_0) \in GL(n)$, as we just defined it; the *linear holonomy group* $\Psi_x$ is then the image of $h$. It is a Lie group and, by the Ambrose-Singer holonomy theorem [**5**], its Lie algebra is generated by all of the possible values of $\Omega^i_j(X,Y)$ when $X$, $Y$ are tangent vectors at $x$. In particular, for a connection with vanishing curvature the Lie algebra of the linear holonomy group is trivial. However, this does not imply that the linear holonomy group of a flat connection is trivial, only that it is discrete – or rather, *totally disconnected*, in the sense that every point of $\Psi_x$ has a neighborhood in $GL(n)$ that contains no other points of $\Psi_x$.

Since all of the linear holonomy groups $\Psi_x$ are isomorphic by conjugation in a path-connected $M$, we denote the generic representative of that isomorphism class by $\Psi$.

One usually thinks of the translational part of the holonomy element as arising from non-vanishing torsion and the linear part from non-vanishing curvature, but we have already seen that both of these geometric objects vanish for the connection $\gamma^i_j$. At first, this appears to lead to an inconsistent conclusion, until we recall that the usual arguments that couple torsion to the translational part of the connection and curvature to the linear part assume that the loop $l_0$ is contractible; more to the point, it must bound a 2-chain: $l_0 = \partial c_2$. One can then apply Stokes's theorem to make:

$$\int_{l_x} \theta^i = \int_{c_2} d\theta^i = 0, \tag{5.7}$$

since we are assuming that the $\theta^i$ are all closed, in order for the integral to be homotopy-invariant.

However, we are considering a multiply-connected $x(\mathcal{O})$, which then implies the existence of non-trivial loops, which do not bound 2-chains, so Stokes's theorem does not apply and the integral on the right-hand side of (5.7) can be non-vanishing. Hence, it is the topology that is responsible for the non-trivial holonomy, not the torsion or curvature.

In order to better see how multiple-connectivity relates to holonomy for flat connections, let us recall some basic properties of flat connections [**5-7**] as they relate to holonomy. One has the following set of equivalent conditions for flatness of a torsion-free linear connection:

**Theorem:**

Suppose $M$ is connected, while $x \in M$, and $\gamma$ is a torsion-free connection on $GL(M)$. The following are equivalent:

1. $\gamma$ is flat.
2. The linear holonomy group $\Psi_x$ is totally disconnected.
3. Parallel translation produces the same element of $GL(n)$ for homotopic loops.
4. The map $\rho : \pi_1(M, x) \to \Psi_x$, $[l] \mapsto h[l]$ is onto.
5. $h = \rho \cdot p$, where $\rho$ is a representation $\rho$: $\pi_1(M, x) \to GL(n)$ and $p$: $\Omega M_x \to \pi_1(M, x)$ takes every loop $l$ to its homotopy class $[l]$.
6. Every homotopically trivial loop $l_x$ makes:

$$0 = \int_{l_x} \gamma^i_j \qquad \left(I = \exp\int_{l_x} \gamma^i_j\right); \tag{5.8}$$

as a consequence, when $M$ is simply-connected and $\gamma$ is flat this is true for any loop.

By contraposition, in order for a flat connection on $M$ to have non-trivial holonomy, $M$ must be multiply connected.

Another important topological property of flat connections on $G$-principal bundles (where $G$ is a Lie group) that relates to simple connectivity is that when the base manifold $M$ of a $G$-principal bundle $P \to M$ is paracompact and simply connected [**6**]:

1. There is a bundle isomorphism $\tau: P \to M \times G$; i.e., $P$ is trivializable.
2. $\gamma = \tau^* \gamma_0$, where $\gamma_0$ is the canonical flat connection on $M \times G \to M$, which makes $T(M)$ the horizontal sub-bundle of $T(M \times G)$.

In particular, the bundle of linear frames on $M$ – for which $G = GL(n)$ – must be trivializable, which then implies that a simply connected manifold can admit a flat connection if and only if it is parallelizable. The connection $\gamma$ then represents the teleparallelism connection.

We conclude by mentioning some other useful properties of flat connections:

1. If $M$ is a compact, connected Riemannian manifold then $\gamma$ is flat iff $\Psi$ is finite.
2. If $M$ is complete and connected and $\gamma$ is flat then $M = A(n)/\pi_1(M, x)$. The value of $h(l)$, when $l(s) = (A^i(s), B^i_j(s))$ is a loop in $A(n)$, will be the linear part $B^i_j(s)$ of $l(s)$ when one expresses $A(n)$ in the form of the semi-direct product $\mathbf{R}^n \times_s GL(n)$.

## 5 Discussion

Let us summarize the essential points of the foregoing discussion:

1. An integrable strain is defined by a diffeomorphism $y: x(\mathcal{O}) \to \bar{x}(\mathcal{O})$ of the initial state $x(\mathcal{O})$ of an object into its deformed state $\bar{x}(\mathcal{O})$, which then defines a holonomic coframe field $dy^i$ on a coordinate neighborhood on $x(\mathcal{O})$.
2. An anholonomic frame field $\theta^i$ on the initial state is therefore not integrable into a diffeomorphism. If the $\theta^i$ are closed then one can, however, define a singular diffeomorphism – viz., an immersion.
3. It is possible to construct this immersion such that the jump discontinuity is a finite affine transformation of the initial point, not an infinitesimal rigid motion.
4. The connection $\gamma^i_j$ that makes $\theta^i$ parallel has vanishing torsion, curvature and non-metricity with respect to that coframe field and the (deformed) metric that makes it orthonormal, but acquires torsion and non-metricity in a holonomic coframe field that is orthonormal for the undeformed metric.
5. When $x(\mathcal{O})$ is multiply-connected, $\gamma^i_j$ can still have non-trivial holonomy around loops that do not contract to points in $x(\mathcal{O})$.

Since the flatness of the connection $\gamma^i_j$ is equivalent to the homotopy-invariance of the holonomy element $h(l) \in A(n)$ that gets associated with the loop $l$, one then sees that in order to be dealing with a connection that has non-vanishing torsion or curvature, one must sacrifice the homotopy invariance of the integration that we have been using to obtain the "singular displacement" from the given strain. Thus, a different choice of loop through an initial point $x_0 \in x(\mathcal{O})$ would produce a different holonomy factor at the same point; similarly, a homotopic loop through a different initial point would produces a different holonomy factor.

One sees that one is dealing with issues in the context of strain compatibility that are common to other established applications of the geometrical and topological methods of mathematical physics, such as topological defects in order media and non-trivial vacua in gauge field theories. The general picture that is emerging seems to be that the link to showing how topological defects can serve as the sources of geometrical fields is undoubtedly based in the integrability of the field equations.